\documentstyle[prl,preprint,aps]{revtex}
\begin{document}
\tighten
\draft
\preprint{TECHNION-PH-94-2}
\title{What is the statistical significance\\ of the solar
neutrino flux problem?}
\author{Jacques Goldberg}
\address{Department of Physics, Technion, Haifa, Israel}
\date{\today}
\maketitle
\begin{abstract}
The discrepancy between measured and predicted neutrino fluxes on
earth may be less significant than generally believed. Indeed,
a technically incorrect method has so far been used to propagate the
errors on the standard solar model input parameters through the
calculation of the predicted neutrino flux. Input parameters should have
been drawn from uniform not Gaussian distributions. The uncertainty on the
flux may therefore
have been substantially underestimated and the significance of
the observed discrepancy correspondingly overestimated. Results from
different experiments which cannot be directly
compared without reference to a model, may also be less incompatible
than currently believed.

\begin{center}
(Submitted to Physical Review Letters)
\end{center}
\end{abstract}


Uncomfortably, an almost trivial remark must be made about the technique
employed to compare the measured flux on Earth of solar neutrini,
with the prediction of the standard solar model.
For a clear statement of the problem,
the reader is referred to John Bahcall's concise talk at the XXVI International
Conference on High Energy Physics\cite{JBDAL}, with apologies to all other
scientists who contributed so much to this subject.

Very briefly, the flux of solar neutrini on Earth
is measured by several experiments using
various techniques\cite{DAVIS,GALLEX,SAGE,KAMIO}. This flux is independently
predicted using the standard solar model\cite{JB88,CT,JB92}. The prediction
depends on a collection of input parameters derived from prior experiments with
substantial uncertainties. These uncertainties are propagated in the neutrino
flux prediction, which is found incompatible with the experiments,
and further renders
incompatible such experiments that cannot be directly compared
but are nevertheless related through the model\cite{JBDAL}.

The complexity of the standard solar model and the subsequent neutrino
flux derivation precludes the straightforward analytical computation of the
covariance matrix which, together with the errors on the input parameters,
would supply the uncertainty associated with a predicted value. Standard
solar model authors\cite{JB88,CT} estimate this uncertainty by repeating
the calculation of the model with several, typically one thousand\cite{JBDAL},
sets of randomly selected input parameters, each drawn from a Gaussian
distribution defined by the experimental
value and error for the input parameter.

This technique is of course perfectly legitimate {\em except that the
sets of input parameters should be drawn from uniform, not Gaussian,
distributions.}

The rest of this note explains why, and what the consequences may be.

The source of this widely spread wrong practice
is the confusion between quantities whose
nature is random and quantities which are fixed even if unknown.

When an experimentalist measures some physical parameter, the
outcome of the measurement is a random variable, but the parameter is
a fixed constant of perfectly well defined value even if nothing is known
yet about that value. Usually, many independent random sources (of finite
variances to be absolutely correct) contribute to the probability distribution
of the outcome of the measurement which is Gaussian by virtue of the central
limit theorem; this Gaussian probability has a very well defined but unknown
expectation value (the actual value of the parameter being measured), and its
variance is the sum of the contributing variances.

When the value of the parameter is known, it is legitimate to
simulate the possible outcome of a measurement by drawing random numbers
from a Gaussian distribution of known expectation value and variance.
This is common practice, for example, in High Energy Physics
detector simulation, where
particle momenta are originally produced by an event generator with strict
momentum conservation, and are then smeared, and a detector momentum
dependent response
guessed, by adding random errors.

Because a physical parameter is not a random number, it is
completely meaningless to talk about the probability that such a parameter
has some given value\cite{SLM}. Usual intuitive speculations about the value
of an unknown parameter
begin by wrongly considering it as a probabilistic quantity. What an
experimental result stated as $P=\hat P\pm\sigma$ teaches us about the
actual value of some parameter $P$ which is being measured, is simply
this\cite{RPP}:

``There is no more than 31.73\% probability to be wrong when asserting that
the interval $[\hat P-\sigma,\hat P+\sigma]$ covers the actual value
of the parameter $P$.''

This is the common
definition of a confidence interval. Nothing is gained, no
statement is made more conservative, by making the interval twice as large
and correspondingly limiting the risk for a wrong statement to 4.55\%, or
any other combination, as long as the error distribution is Gaussian.
There is no additional information about where the actual value of $P$ is
located; it is even not necessarily covered by the confidence interval.
There is therefore no
reason to give preference to any value. After having taken a calculated risk
to be wrong when taking for granted that the actual value is indeed covered
by the confidence interval, there still is no reason to give preference
to any value within that interval.
Therefore, values to be tried for the parameter should be selected uniformly
over the arbitrarily wide confidence interval (without forgetting the risk
of systematic wrong choice since the interval may fail to cover
the actual value of the parameter), rather than following a
Gaussian distribution which very strongly prefers points close to the center
of the interval. The common intuition that values close to the center are
more likely, simply fails to recognize that the unknown but fixed central
value (the parameter) of a Gaussian distribution and one random but known
outcome (the measurement) drawn from this distribution are not functionally
symmetrical and may therefore not be exchanged.

By taking an infinitely wide confidence interval, the probability to make
a wrong statement when saying that this interval covers the actual value of
the parameter vanishes. This is the uninteresting situation in which
all models fit all data. Remembering that $\sigma$ is the initial
experimental error on the input parameter, the width
$2k\sigma$ of the interval to be used can however
be selected {\em after} an arbitrary (reasonable) threshold for wrong
statements $\epsilon$ has been set. For example, if the author of this note
would always
have set $\epsilon$ to the one standard deviation level, 31.73\% of
the papers carrying his name would contain false results. Before invoking
the need for new Physics following a discrepancy between an experiment and
a model, one might probably at least want a safer 1\% level, still
 one or two wrong
papers in a life time, which corresponds to $k=2.58$ ( 2.58
standard deviations).

The extension to more than one parameter is straightforward even if, again,
seemingly anti--intuitive. For a point in $n$-dimensional space to be located
outside of a given hypercube, it is necessary and sufficient that at least
one coordinate be located outside the corresponding confidence interval; more
coordinates out of their confidence intervals are not relevant. The probability
that at least one confidence interval does
not cover the corresponding parameter, which is
simply the complement of the probability that all intervals
cover their own parameter, is $1-(1-\epsilon)^n\approx n\epsilon$. The common
reduced width of uniform distributions for $n$ parameters is therefore the
number of standard deviations at which the Gaussian tail is equal to
$\epsilon/n$ where, we remind, $\epsilon$ is the threshold for the
acceptation of publishing a wrong result.

Predictions for the solar neutrino flux have been compared with
experiment\cite{JBDAL} by generating 1,000 solar models with 1,000
sets of input parameters drawn from Gaussian distributions. Although
finite width uniform distributions will clearly not let the guessed
parameters vary as far as normal distributions, the uniform density in
the hypercube is very different from the clustering around the central
values that Gaussian distributions provide.
Thus, the
distribution of predicted fluxes will almost certainly become
very substantially wider using uniform distributions.
We will however have to wait until the correct
method for guessing parameters is applied
to existing computer programs\cite{JB88,CT},
to see how the significance of
the observed discrepancies is affected.

Discussions with colleagues have shown that the very mundane facts
about statistics quoted above are not well perceived.
A very simple example is therefore presented to illustrate the
issue. Consider a simple pendulum, made of a point mass at the free end
of a thin string of length $l$. The model to be checked says that the
period $T$ is given by the formula $T=2\pi\sqrt{l\over g}$ with
$g={GM\over R^2}$ where $G$ is the gravitational constant, $M$ the mass of
a spherical Earth and $R$ its radius. The length $l$ has supposedly
been directly measured, and
a prediction for $l$ derived from measurements of $\pi,G,M,R,T$. The
measured value of $l$ is compared with the prediction, possibly
requiring ``new Physics''
such as massive strings or non spherical earth or highly viscous air etc...

Let us begin with the physical constant $\pi$, which as everybody knows
has a very well defined value. Hopefully, nobody will ever ask the question
``what is the probability that $\pi=3.17$ ?'', since $\pi$ is not a random
variable. Assume that we get the value of $\pi$ from a very cheap three
digits display pocket calculator, showing $\pi=3.14$. We know nothing about the
next digit (a fresh student might perhaps
write $\pi=3.145\pm0.005$: this however is not fair, because we do not take
31.73\% chances to make a wrong statement when stating that the interval
$[3.14,3.15]$ covers the true value of $\pi$, we take exactly 0\% such risk).
When estimating the effect on the predicted length of the
pendulum of the uncertainty on $\pi$, would it come to anybody's mind to
draw values for $\pi$ from a normal distribution centered around $3.145$
and of $\sigma=0.002$ ($\sigma$ set such that the random number
almost always remains in the $[3.14,3.15]$ interval)? Obviously, one will
rather try each of $0,1,\ldots,9$ in turn as the next digit; random numbers
from an {\em uniform} distribution do the same: the order of the trial is of
no importance, nor the exact value if enough trials are made.

Let us now turn to any of the four other parameters in our trivial problem.
Is there any reason to handle them differently, just because their
100\% confidence
intervals are not finite?

This trivial example has been simulated. Figure \ref{f1}
 shows the expected Gaussian
distribution of a direct measurement of a string of known length, and
simulated model predictions using Gaussian and uniform errors for the input
parameters.
We have artificially
considered $\pi$ as a parameter to enhance the
effect of having many parameters. Values and errors for $G,M,R$ have
been taken from litterature\cite{RPP,HP}, and errors of
the same order of magnitude have been
set for $\pi$ and the ``measured'' length and period. The period was set
to two seconds, the well known beat of a one meter pendulum.
The nominal true length was slightly shifted as shown
in the figure so that Gaussian simulated predictions come out incompatible
within several standard deviations while uniformly simulated ones ``agree''
quite often, always at the 1\% risk of wrong publication level. This exercise
has no other purpose than illustrating the broadening of the distribution
of the model prediction when uniform distributions are used. It does not imply
anything about the solar neutrino flux. At best, it may
hopefully trigger authors
of solar models to rerun their programs with uniform errors.

In conclusion, a more reliable estimate of the significance
of the discrepancy between experiments and standard solar model neutrino
fluxes could easily be obtained by rerunning computer simulations of the
standard solar model with input parameters drawn from uniform rather than
Gaussian distributions, of widths controlled by the priorily set
accepted risk
of publishing a wrong result.

Thanks are due to Arnon Dar and Asher Peres for several illuminating
discussions, and to the students of Physics 114103 whose endless questions
have helped me begin to learn a little bit of statistics.
This work was supported in part by the Technion Fund for the Promotion of
Research.

\begin{figure}
%
\caption{The
expected distributions of direct length measurements (solid line), and
of the length predicted from a measurement
of the period using Gaussian random input parameters (broken line), or
uniform
random input parameters (dotted line). A cross shows the nominal length and
its one standard deviation error.}
\label{f1}
\end{figure}
\end{document}